\documentclass[aps,a4paper,11pt]{revtex4}
\usepackage{graphicx}
\usepackage{amssymb}
\begin{document}

\title{Nanomechanical Properties and Phase Transitions in a Double-Walled (5,5)@(10,10) Carbon Nanotube: {\it ab initio} Calculations}

\author{A.M. Popov$^a$\footnote{ E-mail: am-popov@isan.troitsk.ru}, Yu.E. Lozovik$^a$, A.S. Sobennikov$^b$, A.A. Knizhnik$^c$}

\address{$^a$Institute of Spectroscopy, Troitsk, Moscow region, 142190, Russia\\
$^b$Moscow Institute of Physics and Technology, Dolgoprudnyi, Moscow region, 141701, Russia\\
$^c$Russian Research Centre Kurchatov Institute, Moscow, 123182 Russia}

\begin{abstract}
The structure and elastic properties of (5,5) and (10,10) nanotubes, as well as barriers for relative rotation of the walls and their relative
sliding along the axis in a double-walled (5,5)@(10,10) carbon nanotube, are calculated using the density functional method. The results of these
calculations are the basis for estimating the following physical quantities: shear strengths and diffusion coefficients for relative sliding along
the axis and rotation of the walls, as well as frequencies of relative rotational and translational oscillations of the walls. The
commensurability-incommensurability phase transition is analyzed. The length of the incommensurability defect is estimated on the basis of {\it ab
initio} calculations. It is proposed that (5,5)@(10,10) double-walled carbon nanotube be used as a slider bearing. The possibility of experimental
verification of the results is discussed.
\end{abstract}

\maketitle

\section{INTRODUCTION}

In connection with considerable advances made in recent years in the development of nanomechanics, the search for nanoobjects that can be used as
mobile elements in nanoelectromechanical systems (NEMSs) has become an urgent problem. The possibility of relative motion of the walls
\cite{cumings00,kis06} in multi-walled carbon nanotubes makes it promising to use nanotube walls as mobile elements of NEMSs. At present, nanomotors
in which the walls of a multi-walled nanotube play the role of the shaft and the bush \cite{fennimore03,bourlon04} and memory cells operating on
relative motion of the walls along the nanotube axis \cite{deshpande06} have been achieved experimentally. A number of NEMSs have been proposed,
which operate on the relative motion of carbon nanotube walls, including a gigahertz oscillator \cite{zheng02}, Brownian nanomotor \cite{tu05}, and a
nut-bolt pair \cite{saito01,lozovik03,lozovik03a}. Nanoelectromechanical systems, such as a varying nanoresistor \cite{lozovik03,lozovik03a}, a
stress nanosensor \cite{bichoutskaia06a} and an electromechanical nanothermometer \cite{bichoutskaia07,popov07}, have also been proposed. The
operating principle of these instruments is based on the dependence of the conductivity and the energy of interaction of the walls of a double-walled
carbon nanotube (DWNT) on the relative position of the walls on subnanometer scales. Block diagrams and operating principles of such NEMSs were
considered in \cite{lozovik07}. Thus, analysis of the relative motion and interaction of carbon nanotube walls is very important for developing and
designing NEMSs.

The structure of a nanotube wall is controlled by a pair of integers, chirality indices $(n,m)$ corresponding to the coordinates of the carbon
lattice vector ${\bf c} = n {\bf a}_1 + m {\bf a}_2$, where ${\bf a}_1$ and ${\bf a}_2$ are the unit vectors of the graphite plane. The segment
corresponding to this vector becomes the wall circumference when a fragment of the graphite plane is curled to form the nanotube wall
\cite{class,class1}.

Neither the interaction between carbon nanotube walls nor the interaction between graphite layers have been investigated in details. The results of
measurements and theoretical calculations of the energy of interaction between graphite layers differ by two orders of magnitude (see
\cite{benedict98} and references therein). The shear strengths along the nanotube axis were measured only in a few experiments
\cite{cumings00,kis06}. As regards theoretical studies, the barriers for the relative motion of the walls were calculated for a large number of DWNTs
using semiempirical potentials \cite{saito01,kolmogorov00,damnjanovic02,vukovic03,belikov04}, as well as {\it ab initio} calculations for certain
DWNTs \cite{charlier93,kwon98,palser99,bichoutskaia05,bichoutskaia06}. Barriers for the relative motion of the walls of (5,5)@(10,10) DWNT were
determined using all methods employed for such calculations. The choice of (5,5)@(10,10) DWNTs was dictated by the following considerations: (i) a
small number of atoms in the unit cell of a nanotube, which renders it suitable for {\it ab initio} calculations; (ii) a large height of the barrier
for the relative rotation of the walls can be observed only for DWNTs with compatible rotational symmetries of the walls
\cite{damnjanovic02,belikov04}, including (5,5)@(10,10) DWNT.

Certain semiempirical methods \cite{saito01,belikov04}, as well as {\it ab initio} computational methods \cite{bichoutskaia05,bichoutskaia06} used
for determining the barriers for the relative motion of the walls of (5,5)@(10,10) DWNT, give values of structural parameters and elastic properties
of graphite that are in good agreement with experiment. Nevertheless, the available data for these barriers in Table 1 show that the results of
calculations based on different methods differ by several orders of magnitude. Thus, further development of theoretical methods for studying the
interaction of layers in carbon nanotubes requires that the barriers for the relative motion of the walls be experimentally determined. Comparison of
experimental and calculated heights of these barriers can be used as a criterion for the correctness of the results of calculations.

Here, we will demonstrate that the results of calculating the elastic properties and structure of graphite using the density functional method in the
local density approximation with the basis consisting of a set of plane waves agree well with experiment. This method is used for calculating the
elastic properties, the structure, the energy of interaction, and the height of barriers for the relative motion of the walls in (5,5)@(10,10) DWNT.
The results of these {\it ab initio} calculations are used to estimate the following physical quantities for (5,5)@(10,10) DWNT: shear strengths and
diffusion coefficients for the relative sliding along the axis and relative rotation of DWNT walls, the frequencies of relative rotational and
translational oscillations of the walls, and the energy of formation and length of the incommensurability defect. We consider the possible
experimental study of these quantities.

It should also be noted that the DWNTs under theoretical investigation in this research were obtained using various synthesis methods: in the
standard arc for obtaining carbon nanotubes \cite{ebbesen92}, in the same arc in the presence of hydrogen and a catalyst \cite{hutchison01}, during
catalytic decomposition of hydrocarbons \cite{ci02}, and from single-walled nanotubes with a fullerene chain inside and subjected to heating
\cite{bandow01} or bombardment with electrons \cite{sloan00}.

\section{COMPUTATIONAL TECHNIQUE}

All calculations based on the density functional method in the local density approximation \cite{LDA_ref} with the basis consisting of a set of plane
waves were performed using VASP (Vienna Ab Initio Simulation Package) \cite{Vasp_reference}. The interaction of valence electrons with atomic cores
was described using ultrasoft pseudopotentials \cite{Vasp_PP}. The maximal kinetic energy of plane waves was 358 eV. Our analysis shows that the
error in calculating the energy of the system is less than 0.001 meV/atom for the chosen cutoff energy for the plane waves. In addition, this error
estimate should be insensitive to the structure of the system since the basis of plane waves does not contain an error associated with superposition
of the basis function in contrast to the basis of localized orbitals.

Summation over the Brillouin zone was performed over special points of the Monkhorst-Pack type \cite{Mon_Pack}. The structure and elastic properties
of graphite were calculated using a set of 172 irreducible K points in the Brillouin zone. The barriers for the relative sliding and rotation of the
walls of (5,5)@(10,10) DWNT were analyzed using a set of nine irreducible K points along the axis of nanotubes.

The structure and elastic properties of graphite were calculated as follows. The lengths of vectors $a$ and $c$ of the fundamental translations of
the graphite lattice in the plane of the layers and in the direction perpendicular to this plane varied in the limits 2.39 \AA~ $\leq a \leq$ 2.49
\AA~ and 6.57 \AA~ $\leq c \leq$ 7.90 \AA, respectively. The calculated dependence of total energy $U_g$ of the system on the length of the
fundamental translation vectors of the graphite lattice was approximated by a third-degree polynomial,

{\bf Table 1} Barriers $\Delta U_z$ and $\Delta U_{\phi}$ for relative sliding of the walls along the axis and relative rotation of the walls about
the (5,5)@(10,10) DWNT axis per atom of the outer wall and ratio $\gamma_b=\Delta U_{\phi}/\Delta U_z$ of these barriers

\begin{tabular}{|l|c|c|c|}
\hline Reference & $\Delta U_z$, meV/atom & $\Delta U_{\phi}$, meV/atom & $\gamma_b$ \\
\hline
\cite{saito01}$^a$ & 0.008 & 0.025 & 3.13\\
\cite{belikov04}$^b$ & 0.00745 & 0.0144 & 2.90\\
\cite{vukovic03}$^c$ & 7.5 & 8.7 & 1.16\\
\cite{palser99}$^d$ & 0.128 & 0.438 & 3.47\\
\cite{kwon98}$^d$ & -- & 1.2 & --\\
\cite{charlier93}$^e$ & 0.35 & 0.78 & 2.26\\
\cite{bichoutskaia05}$^f$ & 0.125 & 0.259 & 2.08\\
\cite{bichoutskaia06}$^g$ & 0.138 & 0.407 & 2.85\\
Our data$^e$ & 0.005$\pm$0.001 & 0.205$\pm$0.002 & 41\\
\hline \end{tabular}

\noindent \small{Note: $^a$ Lennard-Jones potential, optimized structure of the walls; $^b$ Lennard-Jones potential, nonoptimized structure of the
walls; $^c$ Crespi-Kolmogorov potential, optimized structure of the walls; $^d$ tight binding method; $^e$ density functional method, local density
approximation, basis set is formed by plane waves; $^f$ density functional method, local density approximation, basis set is formed by pseudowave
$pdpp$ functions; $^g$ density functional method, local density approximation, basis set is formed by pseudowave $dddd$ functions.}

\vspace{1.0cm}

{\bf Table 2} Coefficients of approximation of the dependence of the total energy of graphite on the lengths of fundamental translation vectors of
the lattice by third-degree polynomial (\ref{exp})

 \begin{tabular}{|c|c|c|c|}
            \hline
            Coefficient & Value, eV/unit cell  & Coefficient  & Value, eV/unit cell \\
            \hline
                    &                   &           &           \\
            $D_{00}$    & -40.6169$\pm$0.00004  & $D_{20}$  & -49.0$\pm$0.5     \\
            $D_{30}$    &  0.0580$\pm$0.0003    & $D_{02}$  & 0.036$\pm$0.003   \\
            $D_{03}$    &  46.98$\pm$0.02   & $D_{11}$  & 0.011$\pm$0.033   \\
            $D_{21}$    & -0.069$\pm$0.003  & $D_{12}$      & -0.0223$\pm$0.0002   \\
\hline
\end{tabular}

$$ U_g = D_{00} + D_{30}(c-c_0)^3 + D_{03}(a-a_0)^3 + D_{21}(c-c_0)^2(a-a_0) + D_{12}(c-c_0)(a-a_0)^2 +$$
\begin{equation}\label{exp}
 + D_{20}(c-c_0)^2 + D_{02}(a-a_0)^2 + D_{11}(c-c_0)(a-a_0),
\end{equation}

\noindent where $a_0$ and $c_0$ are the equilibrium lengths of the fundamental translation vectors of the graphite lattice. Table 2 gives the values
of the coefficients obtained as a result of approximation, while Table 3 gives the values of equilibrium lengths of the fundamental translation
vectors of the graphite cell. The elastic constants of graphite were obtained using expression (\ref{exp}) for the total energy of the system,

\begin{equation}
C_{11} + C_{12} = \frac {a_0^2}{2V_0}\frac {\partial^2 U_g}{\partial a^2},~~~~~~~~
C_{33} = \frac {c_0^2}{V_0}\frac {\partial^2 U_g}{\partial c^2},
\end{equation}

\noindent where $V_0$ is the equilibrium volume of the unit cell of graphite. Table 3 contains the elastic constants of graphite. In the same table,
our results of calculating the structural and elastic properties of graphite with the density functional method in the local density approximation
are compared with the results obtained in \cite{Mounet} and the experimental data described in \cite{Mounet}. This comparison shows that the accuracy
of our calculations corresponds to the typical accuracy of calculations by the density functional method.

Thus, our calculation method gives values of the lattice parameters and elastic properties of graphite close to experimental ones. We believe that
this method can also be used for studying the interaction and relative motion of carbon nanotube walls.

\vspace{0.8cm}

{\bf Table 3} Equilibrium lengths of the fundamental translation vectors of the lattice and elastic constants of graphite

 \begin{tabular}{|c|c|c|c|}
 \hline
           & Our result & \cite{Mounet} & Experiment (300K) \\
\hline
$a_0$, \AA  & 2.44140$\pm$0.00001  & 2.440   & 2.4608$\pm$0.0016 \\
$c_0$, \AA  & 6.586$\pm$0.002      &    --     &      --\\
$c_{0}/a_{0}$            & 2.698      & 2.74    & 2.725$\pm$0.001   \\
$C_{11} + C_{12}$, GPa         & 1319   & 1283    & 1240$\pm$40       \\
$C_{33}$, GPa                  & 23.7   & 29      & 36.5$\pm$1        \\
\hline
\end{tabular}

\vspace{0.8cm}
{\bf Table 4} Structural parameters and elastic properties of the (5,5) and (10,10) walls: $R$ is the radius, $t_s$ is the equilibrium
length of a unit cell, and $E_s$ is the Young modulus

\begin{tabular}{|l|c|c|c|c|}
\hline
Wall   & \multicolumn{2}{c|}{(5,5)} & \multicolumn{2}{c|}{(10,10)} \\
\cline{1-5} & our result & available data & our result & available data\\
\hline
$R$, \AA  & 3.40 & 3.408\cite{bichoutskaia06} & 6.77 & 6.748\cite{bichoutskaia06} \\
$t_s$, \AA & 2.439 & 2.4434\cite{bichoutskaia06} & 2.443 & 2.4420\cite{bichoutskaia06}\\
$E_s$, TPa & 1.24 & 0.95\cite{sanchez-portal99}, 0.96\cite{mielke04},
& 1.31 & 0.92\cite{sanchez-portal99}, 0.98\cite{ozaki00}, \\
& & 1.03\cite{bichoutskaia06}, 1.06\cite{vanlier00}& & 1.10\cite{bichoutskaia06}, 1.24\cite{hernandez98}\\
\hline
\end{tabular}

\section{RESULTS AND DISCUSSION}

First, we calculated the equilibrium structure of isolated (5,5) and (10,10) walls. The structure of each wall was optimized for various lengths $t$
of the unit cell of the wall. The dependence of the total energy $U_t$ of the unit cell on its length was interpolated by Hooke law,

\begin{equation}\label{hook}
U_t = U_s + \frac {\mu (t-t_s)^2}{2},
\end{equation}

\noindent where $U_s$ and $t_s$ are the total energy and length of the unit cell, which correspond to the equilibrium structure, and $\mu$ is the
elastic coefficient of a wall of length $t_s$. The values of $U_s$, $t_s$, and $\mu$ were calculated using the least squares method. The Young
modulus of the wall was determined in most publications using the expression

\begin{equation}\label{young}
E_s = \frac{\mu t_s} {\pi R w},
\end{equation}

\noindent where $R$ is the radius of the wall corresponding to the equilibrium structure and $w$ is the effective thickness of the wall: 3.4 \AA
~\cite{bichoutskaia06,vanlier00,sanchez-portal99,mielke04,ozaki00,hernandez98}. Table 4 gives the calculated values of the structural parameters and
the Young modulus of the (5,5) and (10,10) walls, as well as the corresponding data from the literature. It should be noted that the calculated
lengths of the bonds are in conformity with the results of other calculations based on the density functional method \cite{bichoutskaia06}, while the
calculated values of the Young modulus match those calculated using the density functional \cite{bichoutskaia06,sanchez-portal99,mielke04},
Hartree-Fock \cite{vanlier00}, and tight binding methods \cite{ozaki00,hernandez98}.

The difference in the unit cell lengths of the (5,5) and (10,10) walls, $\Delta t=t_2-t_1=0.004$ \AA{}, is three orders of magnitude smaller than the
unit cell lengths. However, a wall with a smaller unit cell length can be extended, while that with a larger unit cell length can contract as a
result of the interaction between the walls \cite{bichoutskaia06}. Here, the energy of interaction between the walls is calculated for DWNTs with
walls having an optimized structure and with a unit cell of the same average length $t_a=(t_2+t_1)/2$.

We determined the energy of interaction between the walls as the difference between the total energy of the DWNT and the total energies of isolated
walls. To analyze the characteristics of the relative motion of nanotube walls, we must calculate the dependence of energy $U$ of interaction between
two adjacent walls on the coordinates describing the relative position of the walls (angle $\phi$ of the relative rotation of the walls about the
nanotube axis and length $z$ of the relative displacement of the walls along this axis).

In accordance with the symmetry of $(n_1,n_1)@(n_2,n_2)$ DWNTs, the unit cell of the potential relief of the interwall interaction energy of such
DWNTs is a rectangle with sides $\delta_z=t_a/2$ and $\delta_\phi=\pi GCD(n_1,n_2)/n_1n_2$, where $GCD(n_1,n_2)$ is the greatest common division of
numbers $n_1$ and $n_2$ \cite{damnjanovic99}. For (5,5)@(10,10) DWNT, we have $\delta_\phi=\pi/10$. The interwall interaction energy $U(z,\phi)$ for
(5,5)@(10,10) DWNT was calculated for 100 relative positions of the walls, corresponding to a unit cell of the potential relief. We found that
potential energy $U_{int}$ of the interaction between layers, which corresponds to the minimum of function $U(z,\phi)$, is 14.5 meV per atom of the
outer wall. This result is in good agreement with the experimental values of the energy of interaction between graphite layers (35$\pm$10 meV/atom)
\cite{benedict98} and between the walls of multi-walled nanotubes with large diameter (20--33 meV/atom) \cite{kis06}, as well as with the results of
calculation by the density functional method in the local density approximation with the basis set of pseudowave $dddd$ functions (23.83 meV/atom of
the outer wall of a (5,5)@(10,10) DWNT) \cite{bichoutskaia06}.

The relative positions of the walls corresponding to a higher symmetry of a DWNT are singular points of interwall interaction energy $U(z,\phi)$
(extrema or saddle points) \cite{damnjanovic99}. For such relative positions of the walls, some of the $U_2$ axes (binary axes) coincide. The $U_2$
axes are perpendicular to the principal DWNT axis and pass through the middles of all bonds and through the centers of all hexagons of the structure
of the wall. According to our calculations, the $U_2$ axes passing through the middles of tilted bonds of the (5,5) wall and the middles of the bonds
lying in the plane perpendicular to the DWNT axis of the (10,10) wall layer coincide for the relative position of the walls corresponding to the
minimum of the energy of interaction between the walls. The same relative position of the walls corresponding to the minimum of energy$U(z,\phi)$ was
obtained using semiempirical calculations \cite{damnjanovic03}. The relative position of the walls corresponding to the energy minimum in our
calculations is shifted along the $z$ axis by a half-period $\delta_z$ relative to the position obtained from calculations based on the density
functional method in the local density approximation with the basis set of pseudowave $dddd$ functions \cite{bichoutskaia06}.

Table 1 (see above) contains the calculated values of barriers $\Delta U_z$ and $\Delta U_{\phi}$ for relative sliding of the walls along the DWNT
axis and relative rotation of the walls (between their relative positions corresponding to the interaction energy minimum), respectively, as well as
ratio $\gamma_b=\Delta U_{\phi}/\Delta U_z$ of these barriers and analogous results from the literature obtained with the help of semiempirical
methods and {\it ab initio} calculations.

In accordance with calculations based on the density functional method \cite{bichoutskaia06} and semiempirical calculations \cite{belikov04}, the
energy of interaction between commensurate nonchiral DWNT walls $(n_1,n_1)@(n_2,n_2)$ and $(n_1,0)@(n_2,0)$ can be interpolated by the following
expression:

\begin{equation}\label{expansion2}
               U(\phi,z)=U_0-\frac{\Delta U_{\phi}}{2} \cos \left(\frac{2\pi}{\delta_{\phi}}
\phi\right)-\frac{\Delta U_z}{2} \cos \left(\frac{2\pi}{\delta_z}
z\right),
\end{equation}

\noindent where $U_0$ is the mean energy of interwall interaction. Henceforth, we will normalize the values of $U_0$, $\Delta U_z$ and $\Delta
U_{\phi}$ to an atom of the outer wall. In accordance with our calculations, the shape of barriers for relative sliding and rotation of the walls can
also be interpolated by cosinusoids to within the computation error; however, the heights of the barriers for sliding or rotation depend on the
relative rotation or shear of walls, respectively, along the DWNT axis. Table 1 contains our results for barrier heights for relative rotations and
sliding of the walls at the instant of their passage through the relative position of the walls corresponding to the minimum of the interwall
interaction energy. However, the barrier height at the instant of passage through the relative position of the walls corresponding to the energy
maximum of interwall interaction is 2\% lower as compared to that for the passage through the minimum, and the barrier height for the sliding is
lower than the computational error (0.001 meV/atom).

The high value of the barrier ratio for relative rotations and sliding of the walls (see Table 1) also indicates a high ratio of threshold forces for
relative rotations and sliding of the walls. Consequently, we believe that this nanotube can be used as a slider bearing.

A telescopic extension of the inner wall gives rise of a capillary force $F_c$ pulling the inner wall back into the outer wall. The average value of
this force is defined as

\begin{equation}\label{forcecap}
\langle F_c\rangle=\left\langle\frac{d U}{d l}\right\rangle=\frac{U_04n_2}{t_a},
\end{equation}

\noindent where $4n_2$ is the number of atoms in the unit cell of the outer wall. The average capillary force for interwall interaction energy values
$U_0$ and length $t_a$ of the unit cell of the nanotubes calculated here is $\langle F_c\rangle=0.380$ nN.

Our values of barriers $\Delta U_z$ and $\Delta U_{\phi}$ for the relative motion of DWNT walls can be used for calculating the values of a number of
physical quantities characterizing the interaction and relative motion of DWNT walls. Expression (\ref{expansion2}) for the dependence of interwall
interaction energy $U(\phi,z)$ on the relative position of the walls defines the threshold static friction forces $F_z$ and $F_{\phi}$ for the
relative sliding of the walls along the nanotube axis and their relative rotation, respectively:

\begin{equation}\label{fx}
F_z = \frac {4n_2 \pi L_{ov}\Delta U_{z} }{\delta_z t_a},
\hspace{1cm} F_{\phi} = \frac {4n_2 \pi L_{ov} \Delta U_{z} } {\delta_{\phi} R_m
t_a},
\end{equation}

\noindent where $L_{ov}$ is the overlap length of the walls and $R_m$ is the radius of the movable wall. The shear strengths for relative sliding of
the walls along the nanotube axis and their relative rotation are defined, respectively, as

\begin{equation}\label{m}
M_z=\frac {F_z} S, \hspace{1cm} M_{\phi}=\frac {F_{\phi}} S,
\end{equation}

\noindent where $S$ is the area of the overlap surface of the walls,

\begin{equation}\label{square}
S=2\pi L_{ov}\left(\frac{R_1+R_2}2\right)=\pi L_{ov}(R_1+R_2),
\end{equation}

\noindent $R_1$ and $R_2$ are the radii of the inner and outer wall, respectively. In accordance with our calculations, the values of shear strengths
for (5,5)@(10,10) DWNT are $M_z=1.0\pm0.2$ MPa and $M_{\phi}=51.5\pm0.5$ MPa, respectively, for the relative sliding of the walls along the nanotube
axis and their relative rotation. In calculating the shear strengths for the relative rotation of the walls, we assumed that the inner wall is
mobile.

The equality of forces $F_c=F_z$ acting on the movable wall controls the minimal overlap length $L_m$ of the walls for which the static friction
force prevents the inner wall from being pulled by the capillary force:

\begin{equation}\label{lm}
L_m=\frac{\delta_z U_0}{\pi \Delta U_z}.
\end{equation}

\noindent According to our calculations, $L_m=110\pm20$ nm for (5,5)@(10,10) DWNT.

To find the frequency of small rotational oscillations and relative vibrations of the walls along the DWNT axis, we approximate expression
(\ref{expansion2}) for the dependence of the interwall interaction energy on their relative position by a parabolic potential well in the vicinity of
the minimum $U_m$ of this dependence,

\begin{equation}\label{expasion2zf}
U(z')=U_m+\frac{k_zz'^2}{2},
\hspace{1cm} U(\phi')=U_m+\frac{k_{\phi}\phi'^2}{2},
\end{equation}

\noindent where $z'$ and $\phi'$ are the coordinate of displacement relative to the energy minimum of interwall interaction. For
$(n_1,n_1)@(n_2,n_2)$ DWNTs, we have

\begin{equation}\label{k}
k_z=\frac{8 n_2 L_{ov} \Delta U_z}{t_a} \left( \frac {2\pi}{t_a} \right)^2, \hspace{1cm} k_{\phi}=\frac{8 n_2 L_{ov} \Delta U_{\phi}}{t_a}\left(
\frac{n_1n_2} {GCD(n_1,n_2)} \right)^2.
\end{equation}

Frequencies $\nu_{\phi}$ and $\nu_z$ of small relative rotational and axial (along the DWNT axis) oscillations of the walls are defined as

\begin{equation}\label{freq}
\nu_z=\frac1{2\pi}\sqrt{\frac{k_z}{M}}, \hspace{1cm} \nu_{\phi}=\frac1{2\pi}\sqrt{\frac{k_{\phi}}{J}},
\end{equation}

\noindent respectively, where $M$ and $J$ are the reduced mass and moment of inertia of the walls.

In the case when only the inner wall of a DWNT is mobile, its moment of inertia is $J_1=M_1 R_1^2$, where $M_1=4n_1m_0L_1/t_a$ is the mass of the
inner wall, $L_1$ is its length, and $m_0$ is the mass of a carbon atom. Analogously, if only the outer wall is mobile, its moment of inertia and
mass are defined as $J_2=M_2 R_2^2$ and $M_2=4n_2m_0L_2/t_a$, where $L_2$ is the length of the outer wall. In the case when both walls are mobile,
the reduced moment of inertia and the mass of the system are given by $J_{12}=J_1J_2/(J_1+J_2)$ and $M_{12}=M_1M_2/(M_1+M_2)$. The latter case can be
encountered, for example, in zero-gravity conditions in a space-based laboratory.

The frequencies of rotational and axial relative oscillations of the walls are independent of the DWNT length when the lengths of both walls are
identical, or when a shorter wall is mobile and the longer wall is fixed. In these cases, the overlap length of the walls coincides with the length
of the mobile wall, and the expressions of the frequencies of relative axial oscillations of the walls assume the form

\begin{equation}\label{frec_z}
\nu_{z,1}=\frac1{t_a}\sqrt{\frac{2n_2\Delta U_{z}}{n_1m_0}},
 \hspace{1cm}
\nu_{z,2}=\frac1{t_a}\sqrt{\frac{2\Delta U_{z}}{m_0}},
\end{equation}
\begin{equation}\label{frec_z1}
\nu_{z,12}=\frac1{t_a}\sqrt{\frac{2(n_1+n_2)\Delta
U_{z}}{n_1m_0}}.
\end{equation}

\noindent Here, $\nu_{z,1}$, $\nu_{z,2}$, and $\nu_{z,12}$ correspond to the inner, outer, and both mobile walls. The expressions for frequencies
$\nu_{\phi,1}$, $\nu_{\phi,2}$, and $\nu_{\phi,12}$ of relative rotational oscillations of the walls corresponding to the inner, outer, and both
mobile walls, respectively, assume the form

\begin{equation}\label{frec_phi1}
\nu_{\phi,1}=\frac{n_2}{\pi GCD(n_1n_2)R_1}\sqrt{\frac{n_1n_2\Delta U_{\phi}}{2m_0}},
 \hspace{1cm}
\nu_{\phi,2}=\frac{n_1n_2}{\pi GCD(n_1n_2)R_2}\sqrt{\frac{\Delta U_{\phi}}{2m_0}},
\end{equation}
\begin{equation}\label{frec_phi12}
\nu_{\phi,12}=\frac{n_2}{\pi
GCD(n_1n_2)R_1R_2}\sqrt{\frac{n_1(n_1R_1^2+n_2R_2^2)\Delta U_{\phi}}{2m_0}}.
\end{equation}

As a result of our ab initio calculations, we obtained the following values of frequencies of relative oscillations of the walls:
$\nu_{z,1}=1.74\pm0.17$, $\nu_{z,2}=1.22\pm0.12$, $\nu_{z,12}=2.1\pm0.2$ $\nu_{\phi,1}=12.66\pm0.06$, $\nu_{\phi,2}=4.50\pm0.02$ and
$\nu_{\phi,12}=13.44\pm0.07$ cm$^{-1}$.

If the energy of thermal motion of a short mobile wall of length $L$ is much lower than the height of the barriers for the sliding of this wall along
the DWNT axis ($k T \ll 4n_2\Delta U_zL/t_a$) and/or for its rotation ($k T \ll 4n_2\Delta U_{\phi}L/t_a$), then hopping diffusion of the shorter
wall along the DWNT axis and/or its hopping rotational diffusion, respectively, takes place \cite{lozovik03,lozovik03a}. In this case, diffusion
coefficients $D_z$ and $D_{\phi}$ for the motion of the mobile wall along the axis and for rotational diffusion are defined as

\begin{equation}
                D_z =  \frac{1}{2} \Omega_z \delta_z^2 \exp \left( - \frac{4n_2\Delta U_zL}{kTt_a} \right)
\end{equation}

\begin{equation}
                D_{\phi} =  \frac{1}{2} \Omega_{\phi} \delta_{\phi}^2 \exp \left( - \frac{4n_2\Delta U_{\phi}L}{kTt_a} \right)
\end{equation}

\noindent respectively, where $\Omega_z$ and $\Omega_{\phi}$ are the preexponential factors in the Arrhenius formula for the frequency of the jump of
the mobile wall between equivalent minima of interwall interaction energy.

Simulation based on the molecular dynamics method shows that the preexponential factor in the Arrhenius formula for the frequency of the jump of a
C$_{60}$@C$_{240}$ nanoparticle shell between equivalent energy minima of interwall interaction is $\Omega=540\pm180$ GHz \cite{lozovik00}. The
frequency of small relative rotations near the minima of interwall interaction energy calculated here for such a nanoparticle is $\nu=60$ GHz for the
same shape of the shells and the same potential of interaction between atoms of the shells. Thus, ratio $\Omega/\nu$ of the preexponential factor in
the Arrhenius formula to the frequency of small oscillations is approximately 10. We believe that ratio $\Omega/\nu$ for other carbon nanostructures
with embedded graphite layers (in particular, carbon nanotubes) is on the same order of magnitude. We used the value of $\Omega/\nu\sim10$ for
estimating the diffusion coefficients of walls in (5,5)@(10,10) DWNT on the basis of our ab initio calculations. The values for the diffusion
coefficients of a short mobile wall of length $L=100$ nm along the DWNT axis at a temperature of 300 K are $D_{z,1}\approx8\cdot10^{-12}$ m$^2$/s and
$D_{z,2}\approx6\cdot10^{-12}$ m$^2$/s for the inner and outer wall, respectively.

Under the action of force $F_d$ directed along the DWNT axis, the shorter wall may drift along the longer one at a velocity $v=B_zF_d$, where $B_z$
is the mobility of the shorter wall for its motion along the DWNT axis. In the case of a small force ($F_d\delta_z \ll 2\Delta U$, where $\Delta U$
is the barrier height for the motion of the shorter wall along the DWNT axis), mobility $B_z$ is related to the diffusion coefficient by the Einstein
relation $D_z=kTB_z$ \cite{lozovik03,lozovik03a}. According to our estimates, for the drift of a wall 100 nm in length at a velocity of 0.1 m/s at a
temperature of $T=300$ K, forces of $F_{d,1}\approx50$ pN and $F_{d,2}\approx70$ pN must be applied to the inner and outer walls, respectively.

Rotational diffusion for walls of length $L=100$ nm at a temperature of 300 K in (5,5)@(10,10) DWNT is ruled out in view of the large height of the
barrier for the relative rotation of the walls. For a mobile wall of length $L=10$ nm, the values of rotational diffusion coefficients at 300 K are
$D_{\phi,1}\approx2\cdot10^4$ rad$^2$/s and $D_{\phi,2}\approx6\cdot10^3$ rad$^2$/s for diffusion of the inner and outer wall, respectively.

A theory of the commensurability-incommensurability phase transition in DWNTs with nearly commensurate walls has been developed recently
\cite{bichoutskaia06a}. In accordance with this theory, such a DWNT can be either in a commensurate phase with identical lengths of the unit cells of
the walls, or in an incommensurate phase, in which long segments with practically commensurate walls alternate with relatively short
incommensurability defects. The phase of the DWNT is controlled by the value of the incommensurability parameter

\begin{equation}\label{h}
h = \frac{2n_2\Delta U_z}{\mu_{12} (\Delta t)^2},
\hspace{1cm}\mu_{12}=\frac {\mu_1 \mu_2}{\mu_1+\mu_2},
\end{equation}

\noindent where $\mu_1$ and $\mu_2$ are the elastic coefficients of the inner and outer wall, respectively, with the length of a DWNT unit cell.

It should be noted that the incommensurability parameter is the ratio of the difference in energy of interwall interaction for the commensurate phase
and a completely incommensurate state to the elastic energy of the DWNT in the commensurate phase (both values of energy are normalized to a DWNT
unit cell). The DWNT is in a commensurate phase if $h > h_c =\pi^2/8$ and in an incommensurate phase if $h < h_c$. In accordance with our
calculations, $h=0.13\pm0.03$ for (5,5)@(10,10) DWNT. Thus, this DWNT is in an incommensurate phase. In this phase, the DWNT has the following
structure: commensurate regions with the same length of the unit cells of the walls alternate with incommensurability defects. In each
incommensurability defect, the number of unit cells in the wall with a shorter equilibrium length of the unit cell is greater by unity than the
corresponding number in the other wall (incommensurability defects are analogous to dislocations in a crystal). The expression for the length of an
incommensurability defect assumes the form \cite{bichoutskaia06a}

\begin{equation}\label{ldef1}
l_d = \frac {t_a} 2 \sqrt{\frac{\mu_{12} t^2_a}{8n_2\Delta U_z}}
\end{equation}

\noindent Our calculations give an estimate of $l_d = 146\pm21$ nm.

\section{CONCLUSIONS}

Let us consider how to experimentally verify our results. At present, the upper boundary of the shear strength for the relative sliding of the walls
relative to the nanotube axis determined with the help of atomic force microscopy is $M_z<0.04$ ÌÏà \cite{kis06}. In most cases, the barriers for the
sliding of adjacent walls (and, hence, the corresponding shear strength) are negligibly small (due to the incompatibility of translational symmetry
of the walls for incommensurate walls \cite{damnjanovic02} and to incompatibility of helical symmetries of the walls for commensurate chiral walls
\cite{kolmogorov00,belikov04,bichoutskaia05}). In experiment \cite{kis06}, the shear strength was measured only for a single pair of walls of a
multi-walled nanotube, which was characterized by a smaller value of this strength as compared to other pairs of adjacent walls. The chirality
indices of the walls were not determined. Thus, we believe that the experimental value of the upper boundary of the shear strength for the relative
sliding of the walls corresponds to incommensurate or commensurate chiral walls. However, the barrier for the relative sliding of adjacent walls may
attain an appreciable value only for commensurate nonchiral walls (in particular, for (5,5)@(10,10) DWNT)
\cite{kolmogorov00,damnjanovic02,vukovic03,belikov04}. The shear strengths calculated here for (5,5)@(10,10) DWNT exceed by several orders of
magnitude the corresponding boundary of this value (0.04 MPa) measured by an atomic force microscope for the relative sliding of adjacent walls with
indeterminate chirality indices. Consequently, the shear strengths for this DWNT can easily be determined experimentally. In our opinion, the minimal
overlap length of the walls in the case of the telescopic extension of the inner wall, for which the static friction force prevents the inner wall
from being pulled by the capillary force, can also be measured using atomic force microscopy.

The DWNT images obtained by transmission electron microscopy demonstrate that the interwall interaction may result in extension of one wall and
compression of the other wall accompanied by the formation of commensurate regions \cite{hashimoto05}. We believe that transmission electron
microscopy can also be used for observing incommensurability defects in DWNTs.

The calculated frequencies of relative oscillations of the walls can be measured using terahertz spectroscopy or Raman spectroscopy. It should be
noted that Raman spectra have recently been obtained for isolated nanotubes \cite{zhang08}. In such measurements, the frequencies of relative
oscillations of the walls should be distinguished from the frequencies of other DWNT modes corresponding to those in isolated walls. In accordance
with our calculations, the modes of (5,5)@(10,10) DWNT with the lowest frequencies, which correspond to the modes existing in isolated walls, have
frequencies exceeding 40 cm$^{-1}$ \cite{dobardzic03,vukovic06}. Thus, the frequencies of relative oscillations of the walls in (5,5)@(10,10) DWNT
correspond to another frequency range as compared to the frequencies of the remaining modes of this DWNT.

Since the barriers for the relative motion of the walls are controlled by the chirality indices of the walls
\cite{saito01,kolmogorov00,damnjanovic02,vukovic03,belikov04,bichoutskaia05,bichoutskaia06}, the measurements of the shear strengths for the relative
motion of the walls and the frequencies of their relative oscillations should be accompanied by determination of the chirality indices of the walls.
It was shown that the chirality indices of both walls of a DWNT can be determined using electron diffractometry \cite{hirahara06}. We believe that
the characteristics of (5,5)@(10,10) DWNT obtained here on the basis of {\it ab initio} calculations of the interwall interaction energy can be
determined by contemporary experimental techniques. Such measurements would facilitate progress in developing theoretical methods for studying the
interwall interaction in nanotubes, as well as in developing NEMSs based on the relative motion of the walls of nanotubes.

\section*{ACKNOWLEDGMENTS}

This work has been partially supported by the Russian Foundation of Basic Research (grants 08-02-00685 and 08-02-90049-Bel)


\begin{thebibliography}{99}
\bibitem{cumings00} J. Cumings and A. Zettl, Science {\bf 289}, 602 (2000).
\bibitem{kis06} A. Kis, K. Jensen, S. Aloni et al., Phys. Rev. Lett. {\bf 97}, 025501 (2006).
\bibitem{fennimore03} A.M. Fennimore, T.D. Yuzvinsky, W.Q. Han et al., Nature {\bf 424}, 408 (2003).
\bibitem{bourlon04} B. Bourlon, D.C. Glatti, L. Forr\'{o} and A. Bachtold, Nano Lett. {\bf 4}, 709 (2004).
\bibitem{deshpande06} V.V. Deshpande, H.-Y. Chiu, H.W. Ch. Postma et al., Nano Lett. {\bf 6}, 1092 (2006).
\bibitem{zheng02} Q. Zheng and Q. Jiang, Phys. Rev. Lett. {\bf 88}, 045503 (2002).
\bibitem{tu05} Z.C. Tu and X. Hu, Phys. Rev., {\bf 72}, 033404 (2005).
\bibitem{saito01} R. Saito, R. Matsuo, T. Kimura et al., Chem. Phys. Lett. {\bf 348}, 187 (2001).
\bibitem{lozovik03} Yu.E. Lozovik, A.V. Minogin and A.M. Popov, Phys. Lett. A {\bf 313}, 112 (2003).
\bibitem{lozovik03a} Yu. E. Lozovik, A. V. Minogin, and A. M. Popov, Pis'ma
Zh. \'{E}ksp. Teor. Fiz. {\bf 77}, 759 (2003) [JETP Lett. {\bf 77}, 631 (2003)].
\bibitem{bichoutskaia06a} E. Bichoutskaia, A.M. Popov, M.I. Heggie and Yu.E. Lozovik, Fullerenes, Nanotubes
and Carbon Nanostructures {\bf 14}, 131 (2006).
\bibitem{bichoutskaia07} E. Bichoutskaia, A.M. Popov, Y.E. Lozovik et al.,
Phys. Lett. A {\bf 366}, 480 (2007).
\bibitem{popov07} A.M. Popov, E. Bichoutskaia, Yu.E. Lozovik and A.S. Kulish, Phys. Stat. Sol. (a) {\bf 204}, 1911~(2007).
\bibitem{lozovik07} Yu. E. Lozovik and A. M. Popov, Usp. Fiz. Nauk {\bf 177},
786 (2007) [Phys.--Usp. {\bf 50}, 749 (2007)].
\bibitem{class} R. Saito, M. Fujita, G. Dresselhaus, M.S. Dresselhaus,
Appl. Phys. Lett., {\bf 60}, 18, 2204 (1992).
\bibitem{class1} R.A. Jishi, M.S. Dresselhaus, G. Dresselhaus, Phys. Rev. B,
{\bf 47}, 24, 16671 (1993).
\bibitem{benedict98} L.X. Benedict, N.G. Chopra, M.L. Cohen et al.,
Chem. Phys. Lett. {\bf 286}, 490 (1998).
\bibitem{kolmogorov00} A.N. Kolmogorov and V.H. Crespi, Phys.
Rev. Lett. {\bf 85}, 4727 (2000).
\bibitem{damnjanovic02} M. Damnjanovi{\'c}, T. Vukovi{\'c} and I. Milo{\v s}evi{\'c}, Eur. Phys. J. B
{\bf 25}, 131 (2002).
\bibitem{vukovic03} T. Vukovi{\'c}, M. Damnjanovi{\'c} and I. Milo{\v s}evi{\'c}, Physica E {\bf 16},
256, (2003).
\bibitem{belikov04} A.V. Belikov, A.G. Nikolaev, Yu.E. Lozovik and A.M. Popov, Chem. Phys. Lett.
{\bf 385}, 72 (2004).
\bibitem{charlier93} J.-C. Charlier and J.P. Michenaud, Phys. Rev. Lett. {\bf 70}, 1858 (1993).
\bibitem{kwon98} Y.K. Kwon and D. Tomanek, Phys. Rev. B {\bf 58}, 16001(R) (1998).
\bibitem{palser99} A.H.R. Palser, Phys. Chem. Chem. Phys. {\bf 1}, 4459 (1999).
\bibitem{bichoutskaia05} E. Bichoutskaia, A.M. Popov, A. El-Barbary et al.,
Phys. Rev. B {\bf 71}, 113403 (2005).
\bibitem{bichoutskaia06} E. Bichoutskaia, A.M. Popov, M.I. Heggie and Yu.E. Lozovik,
Phys. Rev. B {\bf 73}, 045435 (2006).
\bibitem{ebbesen92} T.W. Ebbesen and P.M. Ajayan, Nature {\bf 358}, 220 (1992).
\bibitem{hutchison01} J.L. Hutchison N.A. Kiselev, E.P. Krinichnaya et al., Carbon {\bf
39}, 761 (2001).
\bibitem{ci02} L. Ci, Z. Pao, Z. Zhou, D. Tang et al., Chem. Phys. Lett. {\bf 359}, 63 (2002).
\bibitem{bandow01} S. Bandow, M. Takizawa, K. Hirahara et al.,
Chem. Phys. Lett. {\bf 337}, 48 (2001).
\bibitem{sloan00} J. Sloan, R.E. Dunin-Borkowski, J.L.
Hutchison et al., Chem. Phys. Lett. {\bf 316}, 191 (2000).
\bibitem{LDA_ref} D.M. Ceperley and B.J. Alder, Phys. Rev. Lett. {\bf 45}, 566 (1980).
\bibitem{Vasp_reference} G. Kresse and J. Furthm\"{u}ller, Phys. Rev. B {\bf 54}, 11169 (1996).
\bibitem{Vasp_PP} G. Kresse and J. Hafner, J. Phys.: Condens. Matter {\bf 6}, 8245 (1994).
\bibitem{Mon_Pack} H. J. Monkhorst and J. D. Pack, Phys. Rev. B {\bf 13}, 5188 (1976).
\bibitem{Mounet} N. Mounet and N. Marzari, Phys. Rev. B {\bf 71}, 205214 (2005).
\bibitem{vanlier00} G. Van Lier, G. Van Alsenoy, V. Van Doren and P.
Geerlings, Chem. Phys. Lett. {\bf 326}, 181 (2000).
\bibitem{sanchez-portal99} D. Sanchez-Portal, E. Artacho, J.M. Soler et al., Phys. Rev. B {\bf 59}, 12678 (1999).
\bibitem{mielke04} S.L. Mielke, D. Troya, S. Zhang et al., Chem. Phys. Lett.
{\bf 390}, 413 (2004).
\bibitem{ozaki00} T. Ozaki, Y. Iwasa and T. Mitani, Phys. Rev. Lett. {\bf 84}, 1712 (2000).
\bibitem{hernandez98} E. Hernandez, C. Goze, P. Bernier, and A. Rubio, Phys. Rev. Lett.
{\bf 80}, 4502 (1998).
\bibitem{damnjanovic99} M. Damnjanovi{\'c}, I. Milo{\v s}evi{\'c}, T. Vukovi{\'c}, and R. Sredanovi{\'c}, Phys. Rev. B
{\bf 60}, 2728 (1999).
\bibitem{damnjanovic03} M. Damnjanovi{\'c}, E. Dobardzi{\'c}, I. Milo{\v s}evi{\'c} et al., New J. Phys. {\bf 5}, 148.1
(2003).
\bibitem{lozovik00} Yu.E. Lozovik and A.M. Popov, Chem. Phys. Lett. {\bf 328}, 355 (2000).
\bibitem{hashimoto05} A. Hashimoto, K. Suenaga, K. Urita et al., Phys. Rev. Lett. {\bf 94} 045504 (2005).
\bibitem{zhang08} L. Zhang, Z. Jia, L. Huang et al., J. Phys. Chem. C {\bf 112}, 13893 (2008).
\bibitem{dobardzic03}E. Dobard{\v z}i{\'c}, I. Milo{\v s}evi{\'c}, T. Vukovi{\'c} et al.,
Eur. Phys. J. B {\bf 34}, 409 (2003).
\bibitem{vukovic06} T. Vukovi{\'c}, S. Dmitrovi{\'c}, and E. Dobard{\v z}i{\'c}, Nanotechnology {\bf 17}, 747 (2006).
\bibitem{hirahara06} K. Hirahara, M. Kociak, S. Bandow et al., Phys. Rev. B {\bf 73}, 195420 (2006).
\end{thebibliography}
\end{document}